\documentclass[hyper,11.5pt,letterpaper]{JHEP3}
\usepackage{epsfig}
\usepackage{amsmath}
\usepackage{subfigure}
\usepackage{amssymb}

\newcommand{\be}{\begin{equation}}
\newcommand{\ee}{\end{equation}}
\newcommand{\bea}{\begin{eqnarray}}
\newcommand{\eea}{\end{eqnarray}}

\title{Hierarchies of Susy Splittings in Holographic Gauge Mediation}
\author{W. Fischler and W. Tangarife Garcia\\
Theory Group, Department of Physics and Texas Cosmology Center\\ The University of Texas at Austin,
TX 78712.
\\ E-mail: \email{fischler@physics.utexas.edu, wtang@physics.utexas.edu}}

\abstract{
We generalize to string theory a construction, originally proposed in field theory, that generates a hierarchy in the splittings between members of super-multiplets in various sectors of the theory.
 }

\keywords{Gauge mediation, superpartner splitting, gauge/gravity duality}

\received{???????? ?st, 2010} \accepted{???????? ?th, 1998}
\preprint{
UTTG-07-11\\TCC-009-11}

\begin{document}




\section{\bf Introduction}
 In a previous paper \cite{ft}, we explored how to generate hierarchies in the splittings between superpartners in gauge mediation scenarios (\cite{gmsb} to \cite{gmsb-end}). We considered the cases in which the hierarchy is such that there is an ``almost supersymmetric" sector that is largely decoupled from the visible sector. The discussion in that paper was in a field theoretical context and, from that construction, it was possible to identify candidates for ``invisible" dark matter in addition to the generation of a very flat scalar potential for the inflaton field. 

\par In this paper we extend the study of those hierarchies to the string theory context. More specifically, we explore the generation of hierarchies in the SUSY-breaking mass contributions for different sectors using the holographic gauge mediation proposed in \cite{hgm1} with the techniques  the Klebanov-Strassler gauge/gravity duality \cite{ks1}. In that work, a metastable non-supersymmetric construction is used in a deformed conifold embedding \cite{dkm}. The matter field content is realized by introducing a set of D7-branes. Here, we will analyze the possibilities of introducing additional matter sectors by using two separate stacks of D7-branes. These two sets of D7s can be thought of as the result of splitting an original stack (with a small number of D-branes, for instance, 7). As a result, we obtain a set of sectors with different mass splittings like those in \cite{ft}, although there are some extra features that were not present in the original work \cite{ft}; for example, some massive vector mesons that weakly connect the visible and invisible sectors. 

\par This article is organized as follows: The first section is dedicated to a quick review of the deformed conifold and the metastable non-supersymmetric system. In the second section, we describe the construction of hierarchies in the splitting in the holographic scenario. We close with some conclusions.


\section{\label{}\bf The deformed conifold and SUSY breaking}

The work in \cite{hgm1} realized gauge mediation of supersymmetry breaking by using the deformed conifold geometry. The warped deformed conifold \cite{ks1} is a solution of type IIB SUGRA on $AdS_5\times T^{1,1}$ (where $T^{1,1}\sim S^2\times S^3$ is the base of the conifold), which is dual to the IR limit of the supersymmetric $\mathcal{N}=1$ field theory arising from a set of $N$ D3-branes and $M$ wrapped D5-branes in the strong `t Hooft coupling limit. The gauge group for this theory is $SU(M+N)\times SU(N)$ and the superpotential for the bifundamental fields, $A_{1,2}\sim (\mathbf{N+M},\mathbf{\overline{N}})$ and $B_{1,2}\sim (\overline{\mathbf{N+M}},\mathbf{N})$, is given by  \cite{KW}
\be
W=\epsilon^{ij}\epsilon^{kl}A_i B_k A_j B_l.
\ee  

The singular conifold is parametrized by the holomorphic equation
\be 
\sum_{i=1}^4z_i^2\,=\,0,\quad z_i\,\in\,\mathbb{C}.
\ee
The metric for the background $AdS_5\times T^{1,1}$ may be written as
\be 
ds^2\,=\,h^{-1/2}(r)(-dt^2+dx_1^2+dx_2^2+dx_3^2)\,+h^{1/2}(r)\,(dr^2+r^2\,ds_{T^{1,1}}),
\ee where 
\be 
ds_{T^{1,1}}\,=\,\frac{1}{9}\left(d\psi\,+\,\sum_{i=1}^2{\rm cos}\theta_id\phi_i\right)^2\,+\,\frac{1}{6}\sum_{i=1}^2\left(d\theta_i^2+{\rm sin}^2\theta_id\phi_i^2\right).
\ee 

The warp factor $h(r)$ is obtained from the solution to the trace of the Einstein's equation. The result is 
\be 
h(r)\,=\,\frac{27\pi \alpha'^2}{4r^4}\left(g_sN\,+\,\frac{3}{2\pi}(g_sM)^2\,{\rm ln}(r/r_0)\,+\,\frac{3}{8\pi}(g_sM)^2\right),
\ee where $r_0$ is the UV scale. This was presented by Klebanov and Tseytlin (KT) in \cite{kt}. This solution has a naked singularity at some $r=r_s$ for which $h(r_s)=0$.

In the solution to the SUGRA equations, the 5-form flux acquires a radial logarithmic dependence,
\be
\tilde{F}_5\,=\,dC_4\,+\,B_2 \wedge F_3\,=\,\mathcal{F}_5 + \star \mathcal{F}_5,\,\,\,\,\,\,\,\,\mathcal{F}_5=27\pi \alpha'^2 N_{\rm eff}(r)\,{\rm vol}(T^{1,1}) 
\ee where 
\be 
F_3\,=\,\frac{M \alpha'}{2}\,\omega_3, \,\,\,\,\,\,B_2\,=\,\frac{3 g_s M \alpha'}{2}\,\omega_2\, {\rm ln} (r/r_0) \label{2-form}
\ee and 
\be 
N_{\rm eff}(r) \,=\,N\,+\,\frac{3}{2\pi}g_s M^2 \,{\rm ln}(r/r_0). 
\ee
The forms $\omega_2$ and $\omega_3$ are defined in equation (\ref{forms-def}) in the appendix.

As the $B_2$ flux goes through a period, $N_{\rm eff}(r)$ will decrease $M$ units, which decreases the 5-form $F_5$ by $M$ units as well.  This behavior is matched to the cascade of Seiberg dualities \cite{SD} that occur in the dual gauge theory. One consequence is that, as the theory flows to the IR, the cascade must stop since $N_{\rm eff}$ cannot be negative. In gravity/gauge duality, the radial coordinate defines the RG scale of the dual theory: $\Lambda\,\sim\,r/\alpha'$ \cite{maldacena}.  The two gauge couplings are determined by
\be  
\frac{8\pi^2}{g_1^2}\,+\,\frac{8\pi^2}{g_2^2}\,=\,\frac{2\pi}{g_s}\,+\,{\rm const.}, 
\ee
\be 
\frac{8\pi^2}{g_1^2}\,-\,\frac{8\pi^2}{g_2^2}\,=\,\frac{1}{g_s\,\pi\,\alpha'}\left(\int_{S^2} B_2 \right) - \frac{2\pi}{g_s}\,=\, 6M\,{\rm ln}(r/r_s)\,+\,{\rm const.},
\ee for a vanishing dilaton field$\Phi$. 

When $r$ becomes small, Klebanov and Strassler argued that the above solution (KT) must be modified in order to remove the singularity. Thus, the conifold is replaced by the deformed conifold
\be 
\sum_{i=1}^4\,z_i^2\,=\,\epsilon^2.
\ee
Indeed, they showed that, in the infrared, the field theory is strongly coupled and has a deformed moduli space, which has $M$ independent branches and each has the geometry of a deformed conifold. 

\par For the case $N\,=\,kM-p$, when a set of $p$ anti-D3-branes are added to the above configuration, the system has a metastable non-SUSY vacuum \cite{kpv}.  At the end of the duality cascade (after $k$ steps), there are only the fractional D3-branes and the anti-D3-branes left. The anti-D3-branes experience a potential towards the bottom of the throat but they are not allowed to reach the origin since the conifold has now a tip; thus, they reach a metastable vacuum. At some point, this state will decay to a supersymmetric vacuum. This occurs when the surrounding flux induces them to puff up to a NS5-brane wrapping the $S^2$ on the tip. The NS5-brane is a source for the $H_3$ flux and, as the brane moves to the minimum in the brane/flux potential, it changes the flux by -1 units. This results in a supersymmetric theory with $M-p$ D3-branes. The metastable state is parametrized by the vacuum energy
\be 
\mathcal{S}\,\sim\,\frac{p}{N}\,{\rm exp}\left(-\frac{8\pi\, N}{3g_s\,M^2}\right)\frac{r_0^4}{\alpha'^4}.
\ee

\par DeWolfe, Kachru and Mulligan (DKM) \cite{dkm} found a solution (to first order) for the backreaction of anti-D3-branes in the KT solution, which is the asymptotic UV approximation of the KS solution. In this case, the metric is  
\be  
ds^2=h^{-1/2}(r)\,\eta_{\mu\nu}dx^\mu\,dx^\nu\,+\,h^{1/2}(r)\left(dr^2+r^2\,\sum_{i=1}^2\left((e^{\theta_i})^2+(e^{\phi_i})^2\right)+r^2e^{2b(r)}(e^\psi)^2\right).
\ee where $e^{\theta_i}$, $e^{\phi_i}$ and $e^\psi$ are 1-forms that are defined in equation (\ref{e-forms}).

\par The SUGRA solution in the Einstein frame, as written in the appendix of \cite{hgm1}, is:
\begin{eqnarray}
F_3\,=\,\frac{3\,M}{2}\,\omega_3,\,\,\,\quad \qquad &&\,\,B_2\,=\,k(r)\,\frac{\omega_2}{3}
,\nonumber\\
\tilde{F}_5\,=\,dC_4\,-\,C_2 \wedge H_3,&&\,\,C_4\,=\,\frac{h(r)^{-1}}{g_s} \,d({\rm vol}_{3,1}), \nonumber \\
C_0\,=\,0, \,\,\,\,\,\,\,\,\,\,\,\,\qquad \qquad && \label{sol-dkm}
\end{eqnarray}
\begin{eqnarray}
\frac{r^4}{\alpha'^2}\,h(r) &=& \frac{27\pi}{4}g_s N\,+\,\frac{1}{8}\left(\frac{9}{2}g_sM\right)^2\,+\,\frac{1}{2}\left(\frac{9}{2}g_sM\right)^2{\rm log}\frac{r}{r_0}\nonumber \\
&&+\frac{\alpha'^4}{r^4}\left(\frac{1}{32}27\pi\,g_sN\,+\,\frac{13}{64}\left(\frac{9}{2}g_sM\right)^2\,+\,\frac{1}{4}\left(\frac{9}{2}g_sM\right)^2{\rm log}\frac{r}{r_0}\right)\mathcal{S} ,\\
e^{2b(r)}&=&1+\frac{\alpha'^4}{r^4}\mathcal{S},\\
\frac{1}{\alpha'}k(r)&=&\frac{9}{2}g_s\,M\,{\rm log} \frac{r}{r_0}\,+\,\frac{\alpha'^4}{r^4}\left(\frac{9}{4}\frac{\pi\,N}{M}\,+\,\frac{33}{4}\,g_s\,M\,+\frac{27}{4}\,g_s\,M\,{\rm log} \frac{r}{r_0} \right)\mathcal{S}, \\
\Phi(r)&=& {\rm log}\,g_s \,+\,\frac{\alpha'^4}{r^4}\left(\phi\,-\,3\mathcal{S}\,{\rm log}\frac{r}{r_0}\right),
\end{eqnarray} where the last four equations include the first corrections in $\mathcal{S}$ and $1/r^4$.

\section{\label{HGM} \bf Holographic gauge mediation and splitting hierarchies}

Following the procedure described in \cite{hgm1} for one matter sector, we will use the warped deformed conifold geometry from the previous section to realize gauge mediation to two separate sectors.  The first sector (which we will call ``visible") is a stack of $K$  D7-branes ($K\ll N$) which are holomorphically embedded according to Kuperstein construction \cite{kup} \footnote{There are other possible embeddings of D7s in the KS geometry; see, for example, \cite{PO}.} :
\be 
z_4\,=\,\mu. \label{mu}
\ee
These D7s fill the four space-time dimensions, extend along the radial coordinate of the conifold and wrap a three-cycle of $T^{1,1}$. The matter fields of this sector (the standard model) are placed at the UV cutoff of the geometry, which implies that the SM fields, on the gauge side, are elementary fields and are decoupled from the hidden sector strong interactions. 
\par
\begin{figure}[h]
\begin{center}
\includegraphics[width=7.5cm]{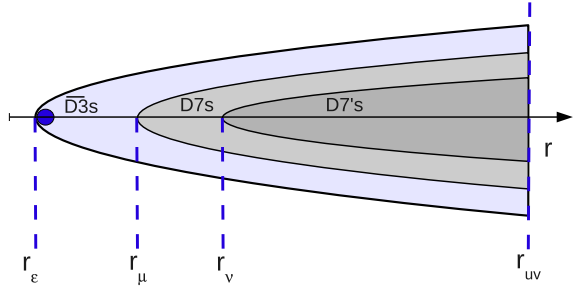}
\end{center}
\caption{Deformed conifold with $p$ ${\rm \overline{D3}s}$ at the tip and two separate stacks of D7s.}
\label{conifold}
\end{figure}

 Likewise, the second sector (``invisible") is a set of $K'$ D7-branes ($K'\,<\,K)$ placed farther than the first stack from the tip of the conifold (see figure \ref{conifold}):
\be 
z_4\,=\,\nu ,\quad \,\nu,\gg\,\mu .  \label{nu}
\ee 

As for the ``visible" sector, the ``invisible" sector can also have a matter content at the UV end of the geometry, from which we can obtain a set of elementary particles that are not coupled to the standard model particles.

\par This theory contains two sets of chiral fields arising from the D7-D3 strings. We have $\chi$ and $\tilde{\chi}$ which transform as $(\mathbf{N},\,\mathbf{\overline{K}})$ and  $(\mathbf{\overline{N}},\,\mathbf{ K})$, respectively. Here, the first element is one of the conifold gauge groups and the second one is the ``visible" sector gauge group. Similarly, we have a second set of chiral fields, $\eta$ and $\tilde{\eta}$, which transform as $(\mathbf{N},\,\mathbf{\overline{K'}})$ and  $(\mathbf{\overline{N}},\, \mathbf{K'})$, under the conifold gauge group and the gauge group of the second sector $SU(K')$ (``invisible" sector). These fields are the messengers of the supersymmetry breaking. The tree-level superpotential is written as 
\begin{eqnarray}
W\,&=&\,\epsilon^{ij}\epsilon^{kl}\,{\rm Tr}(A_i\,B_k\, A_j\, B_l)\,+\,{\rm Tr}(\tilde{\chi}\,A_i\,B_i\,\chi)\,-\,\mu\,{\rm Tr}(\tilde{\chi}\,\chi)+{\rm Tr}(\tilde{\chi}\,\chi\,\tilde{\chi}\,\chi)\nonumber \\
&&+\,{\rm Tr}(\tilde{\eta}\,A_i\,B_i\,\eta)\,-\,\nu\,{\rm Tr}(\tilde{\eta}\,\eta)+{\rm Tr}(\tilde{\eta}\,\eta\,\tilde{\eta}\,\eta). \label{superp}
\end{eqnarray}
The two types of fields, $\chi$ and $\eta$, have dimension $[{\rm mass}]^{3/4}$ and masses $m_\chi=\mu^{2/3}$ and $m_\eta=\nu^{2/3}$, respectively. 
\par We now turn to study the gauge mediation of the SUSY breaking to each of the sectors. There are, in fact, two regimes in which one can compute the soft mass for the gaugino.  The first case is when the mass of the messengers is of the same order as the confining scale of the hidden sector, i.e. $\mu \approx \epsilon$. Thus, the D7s are close to the anti-D3s and we need to use the SUGRA solution close to the tip of the conifold in order to compute the gaugino mass. This approach was studied in \cite{mss} and we will not use it here. 
\par The second regime is the one in which $\mu\gg \epsilon$. This is the scenario proposed in \cite{hgm1} and the one we will use in the rest of this paper. In this case, calculations can be made at large radius on the gravity side (using the KT limit of the KS geometry). Since the messengers are strongly coupled under the conifold gauge group, they form mesons $\Phi$ (or $\Phi'$) which transform in the adjoint representation of the visible (or invisible) gauge group and are weakly coupled. 
\par The masses and $F$-terms for those mesons are computed using holography, where the DBI action for the D7s is dimensional reduced and, as a result, the Kaluza-Klein modes (which are fluctuactions of equations (\ref{mu}) and (\ref{nu})) adquire both supersymmetric and soft masses. The loop interactions of these mesons with the visible (invisible) sector will generate mases for the corresponding gaugino, since $R$-symmetry is broken by the non-zero $F$-term. This sort of gauge mediation, in which the messengers do not participate in the supersymetry breaking, is closely related to ``semi-direct gauge mediation" \cite{sdgm}.  In addition, there are some strings that connect both stacks of D7s. Since we are assuming a large hierarchy between $\mu$ and $\nu$, the states resulting from these strings are quite heavy and the contributions to the masses in either sector are very suppressed. 
\par The bosonic action for the mesons that is obtained from the expansion of the DBI action to quadratic order has the form \footnote{This is a simplified expression that allows us to describe the general form of the soft mass contribution from the mesons. For specific details of the calculation, see the appendices of \cite{hgm1}.}

\begin{eqnarray} 
S\,&=&\,-\frac{\mu_7}{g_s^2}\,\int d^8\sigma\,\left [\mathcal{L}_1\,+\mathcal{L}_2\,\right], \\
\mathcal{L}_1&=&\sqrt{-{\rm det}\gamma}\,\left(g_{4\bar{4}}\gamma^{ab}\partial_a X^4\partial_b X^{\bar{4}}\,+\,g_{44}\gamma^{ab}\partial_a X^4\partial_b X^4\,+\,g_{\bar{4}\bar{4}}\gamma^{ab}\partial_a X^{\bar{4}}\partial_b X^{\bar{4}}\,\right.\nonumber\\&&+\left.\,\gamma^{ac}\gamma^{bd}\,F^1_{cb}\,F^1_{da}\right) \label{action1},\\
\mathcal{L}_2&=&\sqrt{-{\rm det}\gamma'}\, \left(g'_{4\bar{4}}\gamma'^{ab}\partial_a Y^4\partial_b Y^{\bar{4}}\,+\,g'_{44}\gamma'^{ab}\partial_a Y^4\partial_b Y^4\,+\,g'_{\bar{4}\bar{4}}\gamma'^{ab}\partial_a Y^{\bar{4}}\partial_b Y^{\bar{4}}\,\right.\nonumber\\&&+\left. \gamma'^{ac}\gamma'^{bd}\,F^2_{cb}\,F^2_{da}\right), \label{action2}
\end{eqnarray}
where $\mathcal{L}_1$ refers to the stack of D7-branes at $r_\mu$ and $X^4,\,X^{\bar{4}}$ are the small fluctuations about $z_4\,=\mu$. Likewise, $\mathcal{L}_2$ corresponds to the second set of D7s. $\gamma$ and $\gamma'$ are the respective induced metrics on the D-branes and $g,\,g'$ are the metrics of the DKM geometry as defined in equation (\ref{dkm2}) in the appendix.  Here, $\mu_7\,=\,g_s(2\pi)^{-7}$.
\par The next step is to expand $X^4$ and $Y^4$ in Kaluza-Klein modes,
\be 
X^4\,=\,\sum_n\phi_n(x)\xi_n(y),\quad Y^4\,=\,\sum_n\varphi_n(x)\zeta(y),
\ee
where $y$ denotes the internal coordinates of the D7-brane and $x$ denotes the four-dimensional space-time. After plugging these expansions into equations (\ref{action1}, \ref{action2}), we obtain the scalar potential
\be 
V\,=\,\sum_n\left(M_n^2\bar{\phi}_n\phi_n\,+\,F_n \phi_n \phi_n\,+\,{\rm c.c.}\right) \,+\,\sum_n\left(M'^2_n\bar{\varphi}_n \varphi_n\,+\,F'_n \varphi_n \varphi_n\,+\,{\rm c.c.}\right),
\ee
where
\begin{eqnarray}
M_n^2&=&\frac{\int d^4y\,\sqrt{-{\rm det}\gamma}\,g_{4\bar{4}}\gamma^{ab}\,\partial_a\xi(y)\partial_b \bar{\xi}(y)}{\int d^4y\sqrt{{\rm det}\gamma}\,g_{4\bar{4}}\,\xi(y)\bar{\xi}(y)}, \\ F_n&=&\frac{\int d^4y\sqrt{-{\rm det}\gamma}\,g_{44}\gamma^{ab}\,\partial_a\xi(y)\partial_b \xi(y)}{\int d^4y\sqrt{{\rm det}\gamma}\,g_{4\bar{4}}\,\xi(y)\bar{\xi}(y)} ,
\end{eqnarray}
\begin{eqnarray}
M'^2_n&=&\frac{\int d^4y\sqrt{-{\rm det}\,\gamma'}g'_{4\bar{4}}\gamma'^{ab}\,\partial_a\zeta(y)\partial_b \bar{\zeta}(y)}{\int d^4y\sqrt{{\rm det}\gamma'}g'_{4\bar{4}}\,\zeta(y)\bar{\zeta}(y)},\\F'_n&=&\frac{\int d^4y\sqrt{-{\rm det}\gamma'}\,g'_{44}\gamma'^{ab}\,\partial_a\zeta(y)\partial_b \zeta(y)}{\int d^4y\sqrt{-{\rm det}\gamma'}\,g'_{4\bar{4}}\,\zeta(y)\bar{\zeta}(y)}.
\end{eqnarray}
This corresponds, in the superfield formalism, to a superpotential 
\be 
W\,=\,X_n\,\Phi_n\,\Phi_n\,+\,X'_n\,\Phi'_n\,\Phi'_n,
\ee where 
$$\langle\, X_n\rangle\,=\,M_n\,+\,\theta\theta\,F_n\,, \quad \langle\,X'_n\rangle\,=\, M'_n\,+\,\theta\theta\,F'_n\,,$$ and 
$$\Phi_n\,=\,\phi_n\,+\,\theta\psi_n\,+\,\theta\theta\,F_{\phi,\,n},\quad \Phi'_n\,=\,\varphi_n\,+\,\theta\psi'_n\,+\,\theta\theta\,F'_{\phi,\,n}.$$

After introducing the expressions from the DKM metric and integrating over the extra-dimensions, one finds, as in \cite{hgm1},
\begin{eqnarray}
M_n^2\,=\,n^2\frac{|\mu|^{4/3}}{4\pi\,\lambda_{\rm eff}(\mu)}, \quad F_n\,=\,n^2\frac{\bar{\mu}^2\,\mathcal{S}}{|\mu|^{10/3}4\pi\lambda_{\rm eff}(\mu)},\\
M'^2_n\,=\,n^2\frac{|\nu|^{4/3}}{4\pi\,\lambda_{\rm eff}(\nu)}, \quad F'_n\,=\,n^2\frac{\bar{\nu}^2\,\mathcal{S}}{|\mu|^{10/3}4\pi\lambda_{\rm eff}(\nu)}.
\end{eqnarray}
In the last equations, $$\lambda_{\rm eff}(r)\,=\, g_s N_{\rm eff}(r).$$ 

Thus, we notice that the hierarchy between the mass of the messenger mesons is given by 
\be 
\frac{M_n}{M'_n}\,=\,\frac{|\mu|^{2/3}}{|\nu|^{2/3}}\frac{{\rm ln}(r_\mu/r_s)}{{\rm ln}(r_\nu/r_s)} \gg O(1).
\ee

\begin{figure}[h]
\begin{center}
\includegraphics[width=8.5cm]{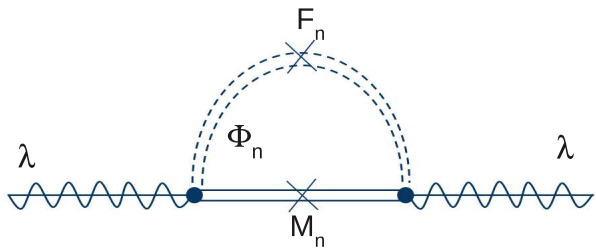}
\end{center}
\caption{1-loop contribution to the gaugino mass from the chiral superfield $\Phi_n$ (the messon messenger).}
\label{loop}
\end{figure}

The gaugino masses for each of the sectors can be computed using field theory. The 1-loop contribution  (figure \ref{loop}) gives
\be 
m^{\rm vis}_\lambda\,=\, \frac{g_{\rm vis}^2\,K}{16\pi^2}\sum_n \frac{F_n}{M_n}\, =\,\frac{g_{\rm vis}^2\,K}{16\pi^2}\frac{\mathcal{S}}{\mu^2\sqrt{4\pi\lambda_{\rm eff}}}\sum_n n e^{i\,\theta_n}
 \ee and 
 
 \be 
m^{\rm inv}_\lambda\,=\, \frac{g_{\rm inv}^2\,K'}{16\pi^2}\sum_n \frac{F'_n}{M'_n}\, =\,\frac{g_{\rm inv}^2\,K'}{16\pi^2}\frac{\mathcal{S}}{\nu^2\sqrt{4\pi\lambda_{\rm eff}}}\sum_n n e^{i\,\theta_n},
 \ee 
 where $\theta_n$ are phases coming from the fact that the $F$-terms are complex. The sum over $n$ is constrained by the binding energy of the mesons, $\mu^{2/3}$ or $ \nu^{2/3}$, which implies $n_{\rm max}<\sqrt{g_s N_{\rm eff}}$. A lower bound for $n_{\rm max}$ is given by the validity of the DBI expansion:
\be 
\sqrt[4]{g_s N_{\rm eff}}<n_{\rm max}<\sqrt{g_s N_{\rm eff}}. 
\ee
 
The authors of \cite{hgm1} noticed that there is a second contribution, due to the deformation in the KS geometry, given by 
\be 
\Delta m\,\sim\, \frac{\epsilon^{2/3}}{|\mu|^{2/3}}\frac{\mathcal{S}}{|\mu|^2\sqrt{4\pi\lambda_{\rm eff}}}
\ee which is much smaller than the contributions from the loop considered above. 

An important remark is that, in order to avoid a Landau pole for the visible and invisible gauge couplings, the geometry must have a cutoff, which implies a constraint for the number of steps in the duality cascade: $k\,<\,5$. The authors of \cite{hgm1} suggested that this constraint might be avoided by working with orbifold geometries.

On the phenomenology aspect of this construction, taking $K\,=\,5$ and $K'\,=\,2$ and $4\pi g_s\,N_{\rm eff}\sim10^2$, we can recreate similar parameters to those in \cite{ft}. For that purpose, we obtain that 
\be 
\mathcal{S}\,\approx\, (10^{10}\,{\rm GeV})^4,\quad \mu\,\approx\, 10^{17} {\rm GeV^{3/2}},
\ee 
 gives a visible gaugino mass of the order of the electroweak scale. For the invisible sector,  if we consider 
\be 
\nu\,\approx\, 10^{20} {\rm GeV^{3/2}},
\ee
we get light ``dark" gauginos ($m^{\rm inv}_\lambda\,\approx\,0.001\, {\rm GeV}$) which may be interesting candidates for ``invisible" components of dark matter. In this context, one could also break the $SU(2)$ gauge symmetry in the invisible sector by separating the two D7-branes which would imply the emergence of invisible massive vector bosons and monopoles.

\section{\bf Conclusions}

We did exhibit in this paper a way to generate hierarchies in the soft masses, for different matter sectors. This was achieved  in string theory using  gauge/gravity duality.  More specifically, we used a warped deformed conifold background with a holomorphic embedding of D7-branes. The hierarchy between the radial position of the two stacks of D7s determined the hierarchy between the messenger masses and, hence, the hierarchy in the SUSY breaking splittings. This kind of hierarchies has phenomenological implications for cosmological models which were studied in previous papers \cite{ft, fm}. It remains important to achieve in string theory, in the context of the construction presented here, a cosmology consistent with observations. 

\section{Acknowledgments}

The research of W.F. and W. T. G. was supported in part by the National Science Foundation under
Grant Numbers PHY-0969020 and PHY-0455649. W. T. G. would like to thank Elena C\' aceres for helpful conversations and the Texas Cosmology Center for partial support.

\appendix
\section{Some definitions on the conifold geometry}

\subsection{The singular conifold}

The unwarped metric on the conifold is given by
\be 
ds_6^2\,=\,dr^2\,+\,r^2ds_{T^{1,1}},
\ee 
\begin{eqnarray} 
ds_{T^{1,1}}&=&\frac{1}{9}\left(d\psi\,+\,\sum_{i=1}^2{\rm cos}\theta_id\phi_i\right)^2\,+\,\frac{1}{6}\sum_{i=1}^2\left(d\theta_i^2+{\rm sin}^2\theta_id\phi_i^2\right) \nonumber \\
&=& \sum_{i=1,2} \left((e^{\theta_i})^2\,+\,(e^{\phi_i})^2 \right)\,+\,(e^{\psi})^2,
\end{eqnarray}
where 

\be 
e^{\theta_i}\,=\,\frac{1}{\sqrt{6}}d\theta_i,\quad e^{\phi_i}\,=\,\frac{1}{\sqrt{6}} {\rm sin}\theta_i\,d\phi_i, \quad e^{\psi}\,=\,\frac{1}{3} \left(d\psi-\sum_{i=1,2} {\rm cos}\theta_i\,d\phi_i \right). \label{e-forms}
\ee

The differential forms in equations (\ref{2-form}, \ref{sol-dkm}) are defined as 
\be 
\omega_2\,=\, 3\left(e^{\theta_1}\wedge e^{\phi_1}\,-\,e^{\theta_2}\wedge e^{\phi_2}\right),\quad \omega_3\,=\, 3e^{\psi}\,\wedge\,\left(e^{\theta_1}\wedge e^{\phi_1}\,-\,e^{\theta_2}\wedge e^{\phi_2}\right). \label{forms-def}
\ee

For more information on the geometric structure of the conifold, see \cite{candelas}.

\subsection{Kuperstein embedding in the DKM solution}

A holomorphic embedding of a set of D7-branes is given by Kuperstein \cite{kup}

\be 
z_4\,=\,\mu.
\ee
The work in \cite{hgm1} used the following parametrization of the holomorphic coordinates of the conifold, where $z_i$ are functions of a new set of coordinates $\rho,\,\chi, \bar{\chi}, \,\gamma, \,\theta,\,\phi$:
\begin{eqnarray}
z_1&=&i(\mu+\chi)\left[{\rm cos} \phi\,{\rm cosh} \left(\frac{\rho+i\gamma}{2}\right){\rm cos}\theta\,-\,i\,{\rm sin}\phi\,{\rm sinh}\left(\frac{\rho+i\gamma}{2}\right)\right] \nonumber \\
z_2 &=&i(\mu+\chi)\left[{\rm sin} \phi\,{\rm cosh} \left(\frac{\rho+i\gamma}{2}\right){\rm cos}\theta\,+\,i\,{\rm cos}\phi\,{\rm sinh}\left(\frac{\rho+i\gamma}{2}\right)\right] \nonumber \\
z_3&=&i(\mu +\chi)\,{\rm cosh}\left(\frac{\rho+i\gamma}{2}\right){\rm sin}\theta\nonumber \\
z_4&=&\mu +\chi,
\end{eqnarray} 
with $\chi \in \mathbb{C}$, $\rho \in [0, \,\infty)$, $\gamma \in [0,\,4 \pi])$, $\theta\in[0,\,\pi]$, $\phi \in [0,\,2\pi)$, and $\chi$, $\bar{\chi}$ are the transverse coordinates. 

With this parametrization, 
\be 
r^3\,=\,\sum_{i=1}^4 |z_i|^2\,=\,|\mu +\chi|^2({\rm cosh}\rho\,+\,1)
\ee
ant the DKM metric is written as 
\begin{equation}
\begin{split} 
ds_6^2 =& \frac{1}{3r}\left\{\frac{|\mu +\chi|^2}{2}\left[ \frac{1+2{\rm cosh}\rho}{3}(d\rho^2+h_3^2)+{\rm cosh}^2\frac{\rho}{2}\,h_1^2+{\rm sinh}\frac{\rho}{2}\,h_2^2\right] \right. \\ &\left. +\frac{2}{3}{\rm sinh}\rho\,(d\rho+i\,h_3)(\mu+\chi)d\bar{\chi}\,+\,{\rm c.c.}\,+\frac{4}{3}(1+{\rm cosh}\rho)d\chi \,d\bar{\chi}\right. \\
& \left. +\frac{\mathcal{S}}{r^4}\left(|\mu +\chi|^2\frac{{\rm cosh}\rho\,-\,1}{3}h_3^2\,+\, \frac{1+{\rm cosh}\rho}{3}\left(2d\chi \,d\bar{\chi}\,-\,\frac{\bar{\mu}\,+\,\bar{\chi}}{\mu +\chi} d\chi^2\,+\,{\rm c.c.}\right) \right. \right. \\ &\left. +\left.\frac{2i}{3}{\rm sinh}\rho \,h_3(\mu \,+\chi)\,d\bar{\chi}\,+\,{\rm c.c.}    \right)\right\}, \label{dkm2}
\end{split}
\end{equation} 
where the $h_i$ forms are defined as 
\be 
\begin{split}
h_1\,=\,& 2\left({\rm cos}\frac{\gamma}{2}d\theta\,-\,{\rm sin}{\gamma}{2}d\phi\right) \\
h_2\,=\,& 2\left({\rm sin}\frac{\gamma}{2}d\theta\,+\,{\rm cos}{\gamma}{2}d\phi\right)\\
h_3\,=\,& d\gamma\,-\,2{\rm cos}\theta\,d\phi.
\end{split}
\ee

The 2-form $B_2$ is written as
\be 
B_2\,=\,\frac{k(r)}{3r^6}|\mu +\chi|^4{\rm cosh}^2\frac{\rho}{2}\left(-{\rm sinh}\frac{\rho}{2}\,d\rho \wedge h_2\,+\,{\rm cosh}\frac{\rho}{2}\,h_3\wedge h_i\right).\ee 
%


\end{document}